\documentclass[aps,twocolumn,showpacs]{revtex4}

\usepackage{amsmath}
\usepackage{graphicx}

\begin{document}
\title[]{Entanglement driven self-organization via a quantum seesaw mechanism}

\author{Christoph Maschler$^{1}$}
\author{Helmut Ritsch$^{1}$}%
\author{Andras Vukics$^{2}$}
\author{Peter Domokos$^{2}$}
\affiliation{$^1$Institut f\"ur theoretische Physik, Universit\"at Innsbruck, Technikerstr.~25, A-6020 Innsbruck,Austria\\
$^2$Research Institute of Solid State Physics and Optics, Hungarian Academy of Sciences, H-1525 Budapest P.O. Box 49, Hungary}

\begin{abstract}
Atom-field entanglement is shown to play a crucial role for the onset of spatial self-organization of ultracold atoms in an optical lattice within a high-{\it Q} cavity. Like particles on a seesaw, the atoms feel a different potential depending on their spatial distribution. The system possesses two stable configurations, where all atoms occupy either only even or only odd lattice sites. While for a classical cavity field description a distribution balanced between even and odd sites is a stationary equilibrium state at zero temperature, the possibility of atom-field entanglement in a quantum field description yields an instant simultaneous decay of the homogeneous atomic cloud into an entangled superposition of the two stable atomic patterns correlated with different cavity fields. This effect could be generic for a wide class of quantum phase transitions, whenever the quantum state can act back on the control parameter.  
\end{abstract}


\maketitle

A spatially modulated laser field far red detuned from an atomic resonance creates a designable optical potential to trap and manipulate ultra-cold atoms~\cite{Kasevich}. For a periodic lattice potential this enables tailored implementations of the Bose-Hubbard Hamiltonian~\cite{jaksch98} to study quantum phase transitions~\cite{bloch02} or ideas of quantum information processing\cite{zoller05}.  While a free-space laser field acts as a prescribed
classical potential, a field enclosed in a cavity is significantly influenced by the atoms and takes part in the coupled atom-field dynamics~\cite{domokos03}.  Nowadays in such setups the regime of strong light-matter
interaction (cavity QED) is experimentally accessible using cold atoms in high finesse Fabry-Perot
resonators~\cite{Rempe,boca04}. Here even a single cavity photon exerts significant forces and the quantum properties of the field can no longer be ignored~\cite{maschler04,vukics05}.

One striking consequence of the coupled atom-field dynamics is spatial self-organization of a laser-illuminated atoms.
Above a threshold pump intensity the atoms spontaneously break the continuous translational symmetry of the cloud and form one of two regular patterns in a phase transition~\cite{domokos02b}. These patterns maximize light scattering from the pump into the cavity with two possible phases of the field, as observed experimentally~\cite{black03}.
The atoms find their stable configurations by a feedback mechanism, i.e:
the optical potential is modified by the cavity field, which is generated by phase coherent scattering of
pump light in the mode by the atoms. Accumulating around every other
antinode the scattering into the cavity mode is enhanced by
constructive interference (superradiance), and the potential depth of those
lattice sites where atoms sit is maximally increased.  

This self-organizing process gets even more intriguing, when the atoms
have a kinetic energy less than the recoil energy, i.e.~their wave
function is flat on the wavelength scale, e.g.\ using a BEC as an
initial state~\cite{Esslinger}. For a ``classical'', or mean-field
description of the light field no self-organization can occur, because of the
destructive interference of the field amplitudes scattered into the
cavity by different parts of the atomic wave function. However, this
conclusion is invalid since the cavity field realizes a {\it quantum
feedback} \cite{lloyd00} for the atomic motion, in which
entanglement is a crucial element.  Scattered field amplitudes with
opposite phases do not cancel but entangle to different atomic wave
functions \cite{cohen}. The quantum average of the field amplitude is still zero,
but the photon number is not, which is clearly incompatible with the
mean-field description.  Field components of the superposition create
different forces, which pull the atomic wave functions towards the
corresponding self-organized configurations. Hence self-organization
is started immediately even at $T=0$ and without measurement induced
projections~\cite{burnett03} (no spontaneous symmetry breaking).

This system is an experimentally accessible implementation of a
``quantum seesaw''. The seesaw is a generic example of a system where
the particle is subject to a dynamically varying potential (feedback). In its
unusual quantum version, the seesaw undergoes an entanglement-assisted
decay from the unstable equilibrium towards the left- and right-tilted
positions.  In this Letter we present a two-site optical lattice model
for the atom-cavity system producing analogous decay. The decay from
an initially symmetric wavefunction is the quantum limit of the
self-organization.  Entangled states, inherent to the effect, are
influenced by cavity losses, thus the lattice model will be
demonstrated by fully quantum Monte Carlo simulations.

\begin{figure}[tp]
\begin{center}  
   \scalebox{0.45}[0.45]{\includegraphics{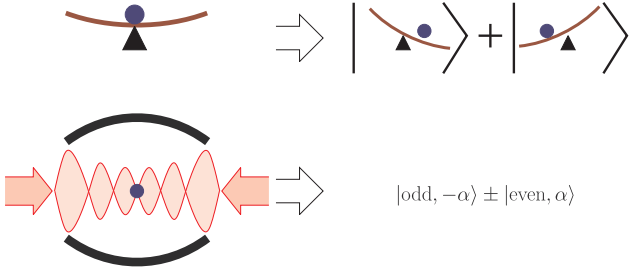}}
  \caption{(color online). Scheme of the quantum seesaw and the corresponding atom-cavity system.}
  \label{fig:scheme}
  \end{center}
\end{figure}

As first toy model we consider a particle moving in 1D along $x$ on a
seesaw potential parameterized by the tilt angle $\varphi$
(Fig.~\ref{fig:scheme}):
\begin{equation}
V(x,\varphi) = \omega_x^2 x^2 + \omega_{\varphi}^2 {\varphi}^2-2J\sin({\varphi})x \; ,
\end{equation}
where harmonic retaining forces have been added both to the particle and
the seesaw (all quantities are dimensionless).  Classically
$x=\varphi=0$ is a stationary point, which is unstable for $J >
\omega_x \omega_{\varphi}$. To model a semiclassical seesaw we
describe $\varphi$ by a classical variable but the $x$-motion quantum
mechanically. Any symmetric wave packet sitting on the balanced seesaw
$\varphi=0$ will exert no net force and is stationary, whereas placing
the center of mass of the wave packet on either side of the seesaw
induces tilting of the potential followed by growing acceleration of
the wave packet.

\begin{figure}[tp]
\begin{center}
   \scalebox{0.42}[0.39]{\includegraphics{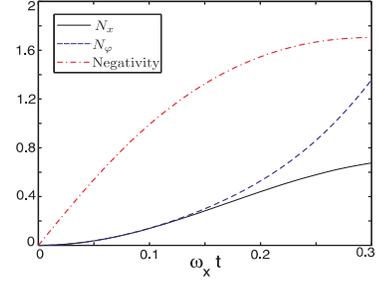}}
  \caption{(color online). Entanglement measured by the negativity and position variance for the system started from the ground state of the uncoupled
    oscillators ($\omega_{\varphi}=3\omega_x $,$J=16/3
    \omega_{\varphi} \omega_x $).}
  \label{fig:time_ev1}
  \end{center}
\end{figure}

The situation is different, however, if $\varphi$ is also a quantum
variable, so that we have a two-component coupled quantum system.
Even for a perfectly symmetric initial condition of the wave packet
and the $\varphi$-oscillator, $\left\langle\varphi\right\rangle=0$,
part of the wave packet will immediately start moving to the right
with growing amplitude $\varphi(t)$, while the second half will move
to the left with opposite tilt $-\varphi(t)$. Thus by forming an
entangled state the wave packet can leave its unstable equilibrium and
escape towards right and left simultaneously. Although the expectation
values of $x$ and $\varphi$ remain zero again, their variances rapidly
grow in time. A simple model Hamiltonian is
\begin{eqnarray}
H&=&\frac{1}{2}({P_x^2}+{P_{\varphi}^2})+V(x,\varphi)\\\nonumber&\approx&
\hbar\omega_x a_x^{\dagger}a_x+\hbar\omega_{\varphi}
a_{\varphi}^{\dagger}a_{\varphi}-\frac{J}{4}(a_{\varphi}^{\dagger}+
a_{\varphi})(a_x^{\dagger}+a_x),
\end{eqnarray}
where the $sin$-function was expanded to first order, valid for small
$\varphi$.  This corresponds to two position-coupled oscillators with
$a_i \;(i\in\{x,\varphi\})$ denoting the annihilation operators.
Starting from the product of the ground states of the uncoupled
oscillators, the system immediately evolves into a strongly entangled
state. The growth of variances and entanglement is shown in
Fig.~\ref{fig:time_ev1}, where this latter is measured by the
negativity of the partial transpose~\cite{Vidal02}.

Let us now turn to ultracold atoms in a 1D optical lattice created by a
standing wave laser field perpendicular to the axis of a cavity
(Fig.~\ref{fig:scheme}). The atoms scatter photons between the cavity
field and the lattice laser, which is associated with momentum
transfer and a modification of the lattice potential. In the limit of
large detuning (small atomic saturation) the single-atom Hamiltonian
reads~\cite{domokos03}
\begin{multline}
\label{eq:Hameff}
H=\frac{p^2}{2m}+ V_0\sin^2(kx)
-\hbar\left(\Delta_c-U_0\right)a^\dagger a
\\+\sqrt{\hbar V_0 U_0}\sin(kx)\left(a+a^\dagger\right)\; ,
\end{multline}
where $a$ and $a^\dagger$ are the cavity photon annihilation and
creation operators. The lattice potential depth is given by $V_0$, the
detuning between the lattice field and the cavity resonance by
$\Delta_c$, and $U_0$ describes the shift of the cavity resonance
frequency per atom.  This enables us to construct an N-atom
Hamiltonian in second quantized form. On expanding the atomic
operators in a localized Wannier basis of the lattice potential and
keeping only the lowest vibrational state, the following Bose-Hubbard
type Hamiltonian~\cite{maschler04} is obtained:
\begin{multline}
\label{BHham}
H=\sum_{m,n}J_{m,n}b_m^\dagger b_n-
\hbar\left(\Delta_c - U_0\sum_n b_n^\dagger b_n\right)a^\dagger a\\+
\left(a+a^\dagger\right)\sum_{m,n}\hbar\tilde{J}_{m,n}b_m^\dagger b_n.
\end{multline}
The nonlinear onsite interaction term is omitted because the atom-atom
collision negligibly contributes to the dynamics. The operator $b_n$
annihilates an atom at the site $kx=(n-1/2)\pi$. The coupling matrix
elements for the kinetic and potential energy $p^2/2m + V_0\sin^2(kx)$
between sites $m$ and $n$ are denoted by $J_{m,n}$, whereas
$\tilde{J}_{m,n}$ gives the matrix elements of
$\sqrt{U_0V_0/\hbar}\sin(kx)$. The model holds as long as the
scattering induced potential change is smaller than the depth of the
lattice potential, $U_0 \ll V_0$.

It is enough to consider only two sites, left and right, centered on
$kx=\pm\pi/2$. The matrix elements $J_{l,r}=J$, $\tilde{J}_{l,l}
= -\tilde{J}_{r,r} = \tilde{J}$ are important, the others either
vanish or amount to additive constants.

\begin{figure}[tp]
  \begin{center}
   \scalebox{0.35}[0.27]{\includegraphics{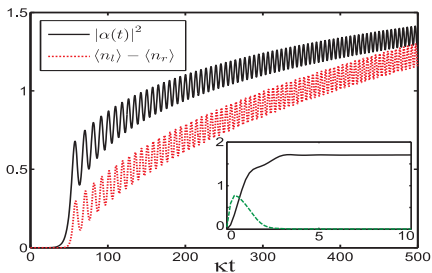}}
     \caption{(color online). (a) Time evolution of the field intensity (dashed line) and site-occupation difference (solid line) for mean-field approximation
       (c.f.~Eq.~\eqref{eq:semicl}) and taking four atoms. The initial
       distribution deviates slightly from a symmetric one,
       $U_0=-0.25\kappa$, $\Delta_c=-2/3\kappa$, $V_0=-4\hbar\kappa$
       and $\kappa=1\frac{\hbar k^2}{m}$.  The insert shows the much
       faster growth of the field intensity (solid line) and
       entanglement of the corresponding quantum model (note the
       different time range).}
  \label{fig:semicl}
  \end{center}
\end{figure}

The ``classical'' mean field approach consists in replacing the field operators $a$ and $a^\dagger$ by their expectation values $\alpha(t)$ and
$\alpha^*(t)$.  The atomic motion is governed by
\begin{equation}
\label{eq:ham_semicl}
H=J\left(b_l^\dagger b_r+b_r^\dagger b_l\right)+
\hbar\tilde{J}\left(b_l^\dagger b_l-b_r^\dagger b_r\right)2\textrm{Re}\left\{\alpha(t)\right\},
\end{equation}
where $\alpha(t)$ fulfills a c-number equation containing expectation
values of atomic operators and a damping term with decay rate
$\kappa$:
\begin{equation}
\label{eq:semicl}
\dot\alpha(t)=
\left[i\left(\Delta_c-U_0N\right)-\kappa\right]\alpha(t)-i\tilde{J}
\left\langle b_l^\dagger b_l-b_r^\dagger b_r\right\rangle\; .
\end{equation}
By construction, entanglement is absent in this model. A perfectly
symmetric atomic distribution with no average field implies
$\dot\alpha(t)=\alpha(t)=0$ and is stationary. As shown in
Fig.~\ref{fig:semicl}, starting with a tiny initial population
asymmetry reveals the bistability of the system and will dynamically
confine the atoms to one of the two wells correlated with a nonzero
field amplitude.
 
\begin{figure}[tp]
\begin{center}
   \scalebox{0.45}[0.42]{\includegraphics{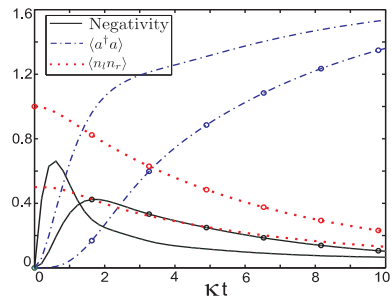}}
  \caption{(color online).  Entanglement (solid lines), photon number (dot-dashed lines),
    and two site atom-atom correlation function (dotted lines) for two
    atoms in two wells. Lines with extra circles show the case of
    exactly one atom in each well , while the others correspond to a
    symmetric superposition state for each atom at start. The
    parameters are $U_0=-2 \kappa, \Delta_c=-6\kappa, J=1/100\kappa$
    and $\tilde{J}=1.6\kappa $}
  \label{fig:time_ev2}
  \end{center}
\end{figure}

In sharp contrast to the mean-field description, the quantum field
model of Eq.~(\ref{BHham}) predicts that the symmetric population
triggers instantly an increase of the photon number associated with
the buildup of atom-field entanglement. This is shown in the insert of
Fig.~\ref{fig:semicl} and also in Fig.~\ref{fig:time_ev2} for two
atoms. This latter plot displays the self-organization process also in
terms of the decaying probability of the two atoms sitting in
different wells ($\langle n_l n_r \rangle$). Note that the
entanglement decays on a longer time scale only after that the
self-organized state with finite intensity was occupied.

We can keep track of the quantum behavior analytically in the bad cavity
limit, where the field is slaved to the atomic motion and can formally
be expressed in terms of the atomic operators:
\begin{equation}
\label{eq:eliminated_field}
a=-i\frac{\tilde{J}}{\kappa -i(\Delta_c - U_0N)}
\left(b_l^\dagger b_l-b_r^\dagger b_r\right).
\end{equation}
As the atom number $\hat{N}=b_l^\dagger b_l+b_r^\dagger b_r$ is a
constant of motion, the dynamics can be consistently restricted to
irreducible subspaces. Let us first consider the single-atom subspace,
where $b_i^\dagger b_i$ acts as a projector on the $i$-th well.
Depending on the atom's position, $\left\langle a\right\rangle$
changes sign, and it vanishes for a symmetric atomic state. However,
the photon number operator $a^{\dagger}a\propto\left(b_l^\dagger
  b_l-b_r^\dagger b_r\right)^2$, which is proportional to the unit
operator in position space. That is, the photon number is nonzero and
independent of the atomic state. 

For more than one atom the dynamics depends also on the quantum
statistics. This is displayed in Fig.~\ref{fig:time_ev2} via the
example of two different quantum states producing the initially
symmetric population distribution. The {\it superfluid} state, when
both atoms are in a symmetric superposition of the two wells,
($1/2 \, (b_l^\dagger+b_r^\dagger)^2 |0\rangle$), self-organizes faster
than the {\it Mott insulator} state ($b_l^\dagger b_r^\dagger
|0\rangle$), indicated by the lines with circles in the figure). The
Mott state is a perfectly balanced initial state with exactly one atom
in each well. It is an eigenstate of the operator $a$, given in Eq.~(\ref{eq:eliminated_field}), with zero eigenvalue, so there is indeed a destructive interference of the quantum mechanical amplitudes of all the excited photon states. Nevertheless, entanglement drives the decay towards the self-organized state via the coherence between the left and right sites induced by tunneling. 



Let us now check the key features of the lattice model by the Monte
Carlo Wave Function (MCWF) method restricted to one
wavelength with periodic boundary conditions \cite{vukics05}. The method allows to
solve the coupled atom-field dynamics given by Eq.~(\ref{eq:Hameff})
including cavity decay and atomic spontaneous emission. Individual
trajectories are simulated and the full atom-cavity density operator
can be approximated by an ensemble average.  This calculation is exempt from constraining the atomic dynamics on the lowest-energy Wannier-basis states. 

\begin{figure}[ht]
  \begin{center}
   \vspace{1cm}
   \scalebox{0.35}[0.27]{\includegraphics{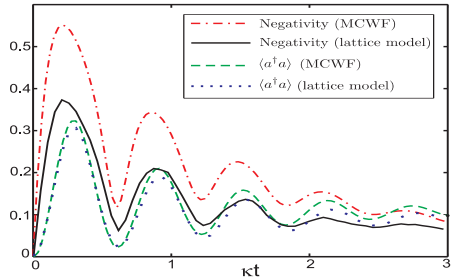}}
  \caption{(color online) Fast dynamic growth of entanglement and photon number of a single atom starting from a flat wavefunction coupled to a vacuum field. Parameters:  $U_0=-1.7\kappa$, $\Delta_c=-12\kappa$, $V_0=-6.7\hbar\kappa$, $\kappa\approx 20\frac{\hbar k^2}{m}$. The results of the MCWF simulations agree well with the lattice model.}
  \label{fig:entanglement}
  \end{center}
\end{figure}

In Fig.~\ref{fig:entanglement} the evolution of the mean photon number and the negativity characterizing the atom-field entanglement is plotted.
Light scattering immediately creates photons entangled to the atomic
state in very good agreement with the quantum lattice model.
Oscillations are related to that the distribution of the number of
spontaneously lost photons is dominated by the 0, 1, 2, etc. numbers.
The photon number oscillates around a constant and does not vanish for
long times. On the other hand, dissipation destroys entanglement so
that the negativity oscillates around a decaying mean. The individual
trajectories reveal that the system evolves into the ``stochastic''
state
\begin{equation}
  \label{eq:Schcat}
 |\psi_\pm\rangle = 1/{\sqrt2}\left(\left\vert\text{left}\right\rangle
    \left\vert\alpha\right\rangle
    \pm\left\vert\text{right}\right\rangle
    \left\vert-\alpha\right\rangle\right),
\end{equation}
where $\left\vert\text{left}\right\rangle$
($\left\vert\text{right}\right\rangle$) means an atomic wave packet
centered on the left (right) site in the lattice potential, and the
radiated field states $\left\vert\alpha\right\rangle$ and
$\left\vert-\alpha\right\rangle$ are coherent ones. The state of
Eq.~\eqref{eq:Schcat} is stochastic in the sense that each photon loss
event flips the sign. After a few jumps the density matrix describes a
left-right mixed state as from the classical model with
random initial conditions.

\begin{figure}[tp]
  \begin{center}
 \scalebox{0.4}[0.32]{\includegraphics[angle=-90]{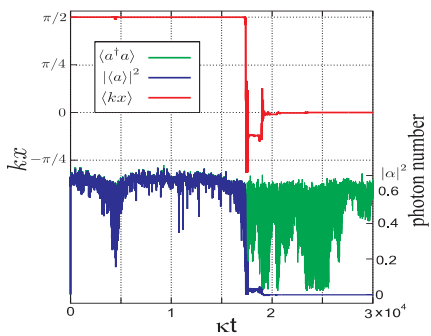}}
  \caption{(color online). Dynamic formation of a symmetrically self-organized state. The atom is
    initially well localized in the right well ($\left\langle x\right\rangle = -\frac\pi 2$) and generates approximately a
    coherent field of intensity $\left\langle a^\dagger
      a\right\rangle\approx\left\vert\left\langle
        a\right\rangle\right\vert^2\approx\left\vert\alpha\right\vert^2$.
    Due to fluctuations it eventually escapes and evolves into the
    state of Eq.~\eqref{eq:Schcat}, having no mean field but non-vanishing photon:
    $\left\langle a \right\rangle=0$, $\left\langle a^\dag
      a\right\rangle\approx\left\vert\alpha\right\vert^2$. Parameters:
    $U_0=-\kappa/2$, $\Delta_c=-1.2\kappa$, $V_0=-2\hbar\kappa$,
    $\kappa\approx 200\frac{\hbar k^2}{m}$.}
  \label{fig:trajectories}
  \end{center}
\end{figure}

Let us start the system from the initial condition
$\left\vert\text{right}\right\rangle$ where it starts to radiate the
coherent field state $\left\vert-\alpha\right\rangle$. As we see in
Fig.~\ref{fig:trajectories}, this state is remarkably stable for a
long time, but eventually collapses due to fluctuations, and the
system ends up in the state of Eq.~\eqref{eq:Schcat}, which is made
obvious by the fact that $\left\langle x\right\rangle=\left\langle
  a\right\rangle=0$, while the photon number is not affected:
$\left\langle a^\dag
  a\right\rangle\approx\left\vert\alpha\right\vert^2$.
Hence in the atom-cavity implementation of the quantum seesaw, even if the seesaw
is tilted to one direction with the atomic wave packet completely on
that side, fluctuations eventually enable the system to escape from this state to
a symmetric final state.

In summary, at the example of self-organization of ultracold atoms in an
optical lattice we found that the possibility of entanglement is an
essential ingredient for dynamical decay of a quantum system from an
unstable equilibrium point. The classical (factorized atom-field
state) description of the optical potential predicts a stationary
homogeneous distribution, while a quantum description implies
immediate atomic ordering via atom-field entanglement formation. The
entanglement involves states, not describable as small quantum
fluctuations around a large mean-field, even if starting the system
from coherent states with large photon and atom number. Entanglement
driven decay can be a generic feature in the dynamics of quantum
phase transitions induced by a classical control parameter, whenever
the quantum system acts even minimally back on its control. Recent
experimental progress in cavity QED should allow to study such models
 with current technology.

We acknowledge funding from the Austrian Science Foundation (P17709),
the National Scientific Fund of Hungary (T043079, T049234), and the
Bolyai Program of the Hungarian Academy of Sciences.


\begin{thebibliography}{99}
\bibitem{Kasevich} M.\ A. Kasevich, {Science\/} {\bf 298}, 1363 (2002).
\bibitem{jaksch98} D.\ Jaksch, C.\ Bruder, J.\ I.\ Cirac, C.\ W.\ Gardiner, and P.\ Zoller, {Phys.\ Rev.\ Lett.\/} {\bf 81}, 3108 (1998); D.\ Jaksch and P.\ Zoller, {Ann.\ Phys.\/} {\bf 315}, 52 (2005).
\bibitem{bloch02}  M.\ Greiner, {\it et.\ al.\/} {Nature\/} {\bf 415}, 39 (2002).
\bibitem{zoller05} P.\ Zoller,  {\it et.\ al.\/},  Eur.\ Phys.\ J.\ D {\bf 36}, 203 (2005) 
\bibitem{domokos03} P.\ Domokos and H.\ Ritsch, {J.\ Opt.\ Soc.\ Am.\ B\/} {\bf 20}, 1098 (2003).
\bibitem{Rempe} P.\ Maunz {\it et.\ al.,\/} {Nature\/} {\bf 428}, 50 (2004).

\bibitem{boca04}A.\ Boca et al, Phys.\ Rev.\ Lett.\ 93, 233603 (2004)
\bibitem{maschler04} C.\ Maschler and H.\ Ritsch, {Phys.\ Rev.\ Lett.\/} (in press); C.\ Maschler and H.\ Ritsch, {Opt.\ Comm.\/} {\bf 243}, 145 (2004).
\bibitem{vukics05}  A.\ Vukics, J.\ Janszky, and P.\ Domokos, J.\ Phys.\ B. {\bf 38}, 1453 (2005)
\bibitem{domokos02b} P.\ Domokos and H.\ Ritsch, {Phys.\ Rev.\ Lett.\/} {\bf 89}, 253003 (2002); J.\ K.\ Asb\'oth, P.\ Domokos, H.\ Ritsch, and A.\ Vukics, Phys.\ Rev.\ A 72, 053417 (2005).
\bibitem{black03} A.\ T.\ Black, H.\ W.\ Chan, and V.\ Vuletic, {Phys.\ Rev.\ Lett.\/} {\bf 91}, 203001 (2003).
\bibitem{Esslinger} A.\ \"{O}ttl, S.\ Ritter, M.\ K\"{o}hl, and T.\ Esslinger, Phys.\ Rev.\ Lett.\ {\bf 95}, 090404 (2005).
\bibitem{lloyd00}S.\ Lloyd, Phys.\ Rev.\ A {\bf 62}, 022108 (2000).
\bibitem{cohen}C.\ Cohen Tannoudji, Atoms in electromagnetic fields, World Scientific Singapore (1994).
\bibitem{burnett03} A.\ V.\ Rau, J.\ A.\ Dunningham, and K.\ Burnett, {Science\/} {\bf 301}, 1081 (2003).
\bibitem{Vidal02} G.\ Vidal and R.\ F.\ Werner, {Phys.\ Rev.\ A\/} {\bf 65}, 032314 (2002).
\end{thebibliography}
\end{document}